\documentclass[12pt]{article}
\usepackage{epsfig}
\usepackage{arydshln}
\date{}
\begin{document}

{\large \sf
\title{
{\normalsize
\begin{flushright}
%arXiv:0709.1526 [hep-ph]
\end{flushright}}
{\vspace{.5cm} {\LARGE \sf Model with Strong
$\gamma_4~T$-violation\thanks{This research was supported in part by
the U.S. Department of Energy (grant no. DE-FG02-92-ER40699)}}
\vspace{1cm}} }

{\large \sf
\author{
{\large \sf
R. Friedberg$^1$ and  T. D. Lee$^{1,~2}$}\\
{\normalsize \it 1. Physics Department, Columbia University}\\
{\normalsize \it New York, NY 10027, U.S.A.}\\
{\normalsize \it 2. China Center of Advanced Science and Technology (CCAST/World Lab.)}\\
{\normalsize \it P.O. Box 8730, Beijing 100080, China}\\
} \maketitle

\begin{abstract}

{\normalsize \sf ~~~~~~

 We extend the $T$ violating model of the
paper on "Hidden symmetry of the CKM and neutrino-mapping matrices"
by assuming its $T$-violating phases $\chi_\uparrow$ and
$\chi_\downarrow$ to be large and the same, with
$\chi=\chi_\uparrow=\chi_\downarrow$. In this case, the model has 9
real parameters:
$\alpha_\uparrow,~\beta_\uparrow,~\xi_\uparrow,~\eta_\uparrow$ for
the $\uparrow$-quark sector,
$\alpha_\downarrow,~\beta_\downarrow,~\xi_\downarrow,~\eta_\downarrow$
for the $\downarrow$ sector and a common $\chi$. We examine whether
these nine parameters are compatible with ten observables: the six
quark masses and the four real parameters that characterize the CKM
matrix (i.e., the Jarlskog invariant ${\cal J}$ and three Eulerian
angles). We find that this is possible only if the $T$violating
phase $\chi$ is large, between $-120^0$ to $-135^0$. In this strong
$T$ violating model, the smallness of the Jarlskog invariant ${\cal
J}\cong 3\times 10^{-5}$ is mainly accounted for by the large heavy
quark masses, with $\frac{m_c}{m_t} <\frac{m_s}{m_b} \approx .02$,
as well as the near complete overlap of $t$ and $b$ quark, with
$(c|b)=-.04$.

\vspace{1cm}
 }
\end{abstract}

{\normalsize \sf PACS{:~~12.15.Ff,~~11.30.Er}}

\vspace{1cm}


\newpage

\section*{\Large \sf 1. Introduction}

In a previous paper on the "Hidden symmetry of the CKM and
neutrino-mapping matrices"[1], we have posited a mass-generating
Hamiltonian $H_\uparrow+H_\downarrow$ where
$$\begin{array}{l}
H_\uparrow = \alpha_\uparrow | q_3^\uparrow - \xi_\uparrow
q_2^\uparrow|^2 + \beta_\uparrow | q_2^\uparrow-\eta_\uparrow
q_1^\uparrow|^2 +\beta_\uparrow |q_3^\uparrow
-\xi_\uparrow\eta_\uparrow q_1^\uparrow|^2 \\
H_\downarrow =
\alpha_\downarrow | q_3^\downarrow - \xi_\downarrow
q_2^\downarrow|^2 + \beta_\downarrow |
q_2^\downarrow-\eta_\downarrow q_1^\downarrow|^2 +\beta_\downarrow
|q_3^\downarrow -\xi_\downarrow\eta_\downarrow q_1^\downarrow|^2
\end{array}\eqno(1.1)
$$
with $\alpha,~\beta,~\xi,~\eta$ real. This conserves $T$ and leads
to zero masses for the light quarks $u$ and $d$. We then modified
(1.1) by replacing $\xi_\uparrow,~\xi_\downarrow$ with the
corresponding $T$ violating factors $\xi_\uparrow
e^{i\chi_\uparrow}$ and $\xi_\downarrow e^{i\chi_\downarrow}$. To
first order in $\chi_\uparrow$ and $\chi_\downarrow$ we found a
relation of proportionality between ${\cal J}$, the Jarlskog
invariant measuring $T$-violation, and a linear combination of
square roots of the light masses. The ratio agreed roughly with
known values. We shall call this the "weak $\gamma_4$-model"
because to make the calculation we assumed
$\chi_\uparrow,~\chi_\downarrow$ to be small.

There were two reasons for dissatisfaction with this model. First,
why not introduce the phase factor into $\eta$ or $\xi\eta$,
yielding different physics? And second, when we estimated not only
${\cal J}$ but the individual matrix elements of $U_{CKM}$, we found
that the data required $\chi_\uparrow$ and $\chi_\downarrow$ to be
large angles, not small.

We now present a new model, the "strong $\gamma_4$-model". In this
model we introduce phase factors independently into all three terms,
but require them to have the same values in $H_\uparrow$ and
$H_\downarrow$. Thus we take the mass-generating Hamiltonian to be
$H_\uparrow+H_\downarrow$ where
$$
\begin{array}{l}
H_\uparrow = \alpha_\uparrow | q_3^\uparrow - \xi_\uparrow e^{i\rho}
q_2^\uparrow|^2 + \beta_\uparrow | q_2^\uparrow-\eta_\uparrow
e^{i\omega} q_1^\uparrow|^2 +\beta_\uparrow |q_3^\uparrow
-\xi_\uparrow\eta_\uparrow e^{-i\tau} q_1^\uparrow|^2 \\
H_\downarrow = \alpha_\downarrow | q_3^\downarrow - \xi_\downarrow
e^{i\rho} q_2^\downarrow|^2 + \beta_\downarrow |
q_2^\downarrow-\eta_\downarrow e^{i\omega} q_1^\downarrow|^2
+\beta_\downarrow |q_3^\downarrow -\xi_\downarrow\eta_\downarrow
e^{-i\tau} q_1^\downarrow|^2
\end{array}.\eqno(1.2)
$$
It is now easily seen that the masses and CKM matrix depend on the
phases only through the sum $\chi=\rho+\omega+\tau$. Accordingly,
without loss of generality, we set $\rho=\omega=0$, $\tau=\chi$. The
mass-generating Hamiltonian can then be written as
$$
\bigg(\bar{q}_1^\uparrow,~
\bar{q}_2^\uparrow,~\bar{q}_3^\uparrow\bigg) M_\uparrow \left(
\begin{array}{r}
q_1^\uparrow\\
q_2^\uparrow\\
q_3^\uparrow
\end{array}\right) +
\bigg(\bar{q}_1^\downarrow,~
\bar{q}_2^\downarrow,~\bar{q}_3^\downarrow\bigg) M_\downarrow \left(
\begin{array}{r}
q_1^\downarrow\\
q_2^\downarrow\\
q_3^\downarrow
\end{array}\right)
$$
where $q_i^\uparrow,~q_i^\downarrow$ and $\bar{q}_i^\uparrow,
~\bar{q}_i^\downarrow$ are related to the corresponding Dirac
field operators $\psi(q_i(\uparrow)),~\psi(q_i(\downarrow))$ and
their hermitian conjugate $\psi^\dag(q_i(\uparrow)),
~\psi~\dag(q_i(\downarrow))$ by
$$
q_i^{\uparrow/\downarrow} =\psi(q_i(\uparrow/\downarrow))~~{\sf
and}~~ \bar{q}_i^{\uparrow/\downarrow}
=\psi^\dag(q_i(\uparrow/\downarrow))\gamma_4,\eqno(1.3)
$$
$$
M_{\uparrow/\downarrow}=\left(
\begin{array}{ccc}
\beta \eta^2(1+\xi^2)& -\beta \eta & -\beta \xi \eta e^{i\chi}\\
-\beta \eta &\beta + \alpha \xi^2 & -\alpha \xi\\
-\beta \xi \eta e^{-i\chi}&-\alpha \xi& \alpha +\beta
\end{array}\right)_{\uparrow/\downarrow}, \eqno(1.4)
$$
with the arrow-subscripts $\uparrow,~\downarrow$ referring to
$\alpha,~\beta,~\xi,~\eta$, but not to $\chi$.

In diagonalizing (1.4) we do not assume, as in the weak
$\gamma_4$-model, that $\chi$ is small. We find that the smallness
of ${\cal J}$ is mainly accounted for by the large heavy masses
with
$$
\frac{m_c}{m_t}<\frac{m_s}{m_b}\approx .02\eqno(1.5)
$$
and by the nearly complete overlap of the statevectors for $t$ and
$b$ since
$$
|(u|b)|<|(c|b)|\cong 0.04.\eqno(1.6)
$$
We have been able to carry out complete calculations in which the
only approximations are based on the smallness of
$\frac{m_s}{m_b},~\frac{m_c}{m_t}$ and $(c|b)$. These calculations
are described in Sections 2 and 3; we give here a brief outline.

We diagonalize $M_\uparrow$ and $M_\downarrow$ with the aid of
parameters $r_{\uparrow,\downarrow},~B_{\uparrow,\downarrow},
~\Phi_{\uparrow,\downarrow},~{\cal S},~{\cal L}$ to be defined in
the next two sections. They are shown there to satisfy the
following ten equations (to first order in small quantities):
$$
\frac{1-r_\uparrow^2}{r_\uparrow^2}\sin^2
B_\uparrow=\frac{4m_um_c}{(m_c-m_u)^2},\eqno(1.7)
$$
$$
\frac{1-r_\downarrow^2}{r_\downarrow^2}\sin^2
B_\downarrow=\frac{4m_dm_s}{(m_s-m_d)^2},\eqno(1.8)
$$
$$
\sin^2\frac{1}{2}\chi=\frac{1-r_\uparrow^2}{\sin^22\Phi_\uparrow}=
\frac{1-r_\downarrow^2}{\sin^22\Phi_\downarrow},\eqno(1.9)
$$
$$
{\cal
L}=\frac{\sqrt{m_sm_d}}{m_b}-\frac{\sqrt{m_cm_u}}{m_t},\eqno(1.10)
$$
$$
{\cal S}=\sin(\Phi_\uparrow-\Phi_\downarrow)=(c|b),\eqno(1.11)
$$
$$
|(u|b)+{\cal S}\sin\frac{1}{2}B_\uparrow|^2={\cal
L}^2\cos^2\frac{1}{2}B_\uparrow,\eqno(1.12)
$$
$$
Im(u|b)=-{\cal L}\frac{\cos
\frac{1}{2}B_\uparrow\cos\frac{1}{2}\chi}{r_\uparrow}\eqno(1.13)
$$
and
$$
(u|s)=\sin\frac{1}{2}(B_\downarrow-B_\uparrow).\eqno(1.14)
$$

Our strategy of solution is as follows. We take
$m_s,~m_c,~m_b,~m_t$, as well as $(u|s),~(u|b)$ and $(c|b)$, to be
given from data (see table 1). Then we have eleven unknowns
$r_{\uparrow,\downarrow},~B_{\uparrow,\downarrow},
~,\Phi_{\uparrow,\downarrow},~{\cal S},~{\cal L},~\chi,~m_d,~m_u$
constrained by ten independent equations given above (with
(1.9)and (1.11), each counted as two equations). Taking a trial
value of $\sin\frac{1}{2}B_\uparrow$, we are able to solve
numerically for the other ten unknowns by a self-correcting double
iteration that converges to 4 decimal stability after $36=6\times
6$ passes. We find that $m_u$ is particularly sensitive to
variations in $\sin\frac{1}{2}B_\uparrow$; a variation of $30\%$
in the latter carries $m_u$ through the whole of its experimental
range from $1.5$ to $3.0MeV/c^2$. Meanwhile $m_d$ varies by only
$25\%$, from $5.2$ to $6.5MeV/c^2$, well within the experimental
range, $3.0$ to $8.0MeV/c^2$. The value of $\chi$ must be taken as
negative and is in the neighborhood of $-125^0$, between $-120^0$
and $-135^0$. We have also tried deviations in
$m_s,~m_b,~(c|b),~Re(u|b)$ and $Im(u|b)$. Only in the case of
$m_s$ does it appear that a maximal deviation $(-25\%)$ from the
"best value" might push $m_d$ outside the range given by data.
(See Tables~1 and 2, and Fig.~1).

The next two sections are devoted to defining the parameters that
appear in (1.7)-(1.14) and proving that these equations are
satisfied. In Section~2, we discuss the separate diagonalization
of $M_\uparrow$ and $M_\downarrow$, and in Section~3, we examine
the CKM matrix.

In Section~4, we discuss briefly a third model[2], which we may
call a $i\gamma_5$ model, because its Hamiltonian contains a term
in $i\gamma_4\gamma_5$ as well as the usual one in $\gamma_4$.

\newpage

\section*{\Large \sf 2. Diagonalization of $M_\uparrow$ and $M_\downarrow$}

In this section, we shall drop the arrow-subscripts and write
(1.4) as
$$
M=\left(
\begin{array}{ccc}
T^2\beta& -T\beta \cos\Phi & -T\beta \sin\Phi e^{i\chi}\\
-T\beta \cos\Phi & \alpha \tan^2\Phi+\beta & -\alpha \tan\Phi\\
-T\beta \sin\Phi e^{-i\chi}&-\alpha \tan\Phi& \alpha +\beta
\end{array}\right), \eqno(2.1)
$$
where
$$
\Phi=\tan^{-1}\xi\eqno(2.2)
$$
$$
T=\eta\sqrt{1+\xi^2}\eqno(2.3)
$$
so that $T^2\beta=\beta\eta^2(1+\xi^2)$,
$\sin\Phi=\xi/\sqrt{1+\xi^2}$, $\cos\Phi=1/\sqrt{1+\xi^2}$ and
(2.1)=(1.4). We denote the eigenvalues of $M$ by $m_l,~m_m,~m_h$
(light, medium, heavy), and seek a unitary matrix $W$ (with
$WW^\dag =1$) such that
$$
M=W\left(
\begin{array}{ccc}
m_l&0&0\\
0&m_m&0\\
0&0&m_h
\end{array}\right)W^\dag.\eqno(2.4)
$$
The $W$ matrix will be built up in stages, as we shall discuss.
First we isolate the heavy mass by writing
$$
M=\Omega\left(
\begin{array}{c;{2pt/2pt}c}
(n)&\begin{array}{c}
L\\
0
\end{array}\\\hdashline[2pt/2pt]
L^*~~~0&\mu_h
\end{array}\right)\Omega^\dag\eqno(2.5)
$$
where

$$
\Omega^\dag=\left(
\begin{array}{c;{2pt/2pt}c}
1&0~~~0\\\hdashline[2pt/2pt]
\begin{array}{c}
0\\
0
\end{array}&e^{i\Phi \tau_y}
\end{array}\right),\eqno(2.6)
$$
$$
\mu_h=\alpha\sec^2\Phi+\beta\eqno(2.7)
$$
$$
L=T\beta\cos\Phi\sin\Phi(1-e^{i\chi})\eqno(2.8)
$$
and
$$
(n)=\beta\left(
\begin{array}{cc}
T^2&-T(\cos^2\Phi+\sin^2\Phi e^{i\chi})\\
-T(\cos^2\Phi+\sin^2\Phi e^{-i\chi})&1
\end{array}\right).\eqno(2.9)
$$
Thus, (2.1) can be obtained by a simple substitution of
(2.6)-(2.9) into (2.5).

Next, we diagonalize the $2\times 2$ matrix $(n)$ of (2.9) by
setting
$$
\cos^2\Phi+\sin^2\Phi e^{i\chi}=r e^{iA}\eqno(2.10)
$$
with $r,~A$ both real. Then
$$
(n)=\beta\left(
\begin{array}{cc}
T^2&-Tr e^{iA}\\
-Tr e^{-iA}&1
\end{array}\right)
$$
$$
=e^{\frac{1}{2}i\tau_zA}e^{-\frac{1}{2}i\tau_yB}\left(
\begin{array}{cc}
\mu_l&0\\
&\mu_m
\end{array}\right)e^{\frac{1}{2}i\tau_yB}
e^{-\frac{1}{2}i\tau_zA}\eqno(2.11)
$$
provided that
$$
\mu_m+\mu_l=\beta(1+T^2)
$$
$$
(\mu_m-\mu_l)\cos B=\beta(1-T^2)\eqno(2.12)
$$
$$
(\mu_m-\mu_l)\sin B=2\beta Tr.
$$
By quadratic combination of (2.12) we obtain
$$
\mu_m\mu_l=\beta^2T^2(1-r^2);\eqno(2.13)
$$
then, by dividing the above equation by the square of the last line
of (2.12), we have
$$
\frac{4\mu_m\mu_l}{(\mu_m-\mu_l)^2}=
\frac{1-r^2}{r^2}\sin^2B\eqno(2.14)
$$
which leads to (1.7) and (1.8).

Also, by applying the Law of Sines to the complex triangle
described by (2.10), followed by trigonometric identities, we find
$$
\cos(\frac{1}{2}\chi-A)=\frac{\cos\frac{1}{2}\chi}{r},\eqno(2.15)
$$
a relation that will be useful later.

Applying (2.11) to (2.5), we now have
$$
M=\Omega V\left(
\begin{array}{ccc}
\mu_l&0&L\Delta^*\cos\frac{1}{2}B\\
0&\mu_m&-L\Delta^*\sin\frac{1}{2}B\\
L^*\Delta\cos\frac{1}{2}B&-L^*\Delta\sin\frac{1}{2}B&\mu_h
\end{array}\right)V^\dag\Omega^\dag,\eqno(2.16)
$$
where
$$
\Delta=e^{\frac{1}{2}iA}\eqno(2.17)
$$
and
$$
V^\dag=\left(
\begin{array}{c;{2pt/2pt}c}
\bigg(e^{\frac{1}{2}i\tau_yB}e^{-\frac{1}{2}i\tau_zA}\bigg)&
\begin{array}{c}
0\\
0
\end{array}\\\hdashline[2pt/2pt]
0~~~~~~~~0&1
\end{array}\right)\eqno(2.18)
$$
Thus $M$ is almost diagonalized. Let us study the magnitude of
$L$. From (2.13) and (2.10) we find
$$
\mu_m\mu_l=\beta^2T^2(1-r^2)=2\beta^2T^2
(1-\cos\chi)\cos^2\Phi\sin^2\Phi\eqno(2.19)
$$
and comparing this with (2.8) we have
$$
|L|=2|T\beta\cos\Phi\sin\Phi\sin\frac{1}{2}\chi|
=\sqrt{\mu_m\mu_l}.\eqno(2.20)
$$
Hence, if we write
$$
\left(
\begin{array}{ccc}
\mu_l&0&L\Delta^*\cos\frac{1}{2}B\\
0&\mu_m&-L\Delta^*\sin\frac{1}{2}B\\
L^*\Delta\cos\frac{1}{2}B&-L^*\Delta\sin\frac{1}{2}B&\mu_h
\end{array}\right)=P\left(
\begin{array}{ccc}
m_l&0&0\\
0&m_m&0\\
0&0&m_h
\end{array}\right)P^\dag\eqno(2.21)
$$
the elements of $P$ will differ from those of the unit matrix by
$O\bigg[\frac{\sqrt{m_lm_m}}{m_h}\bigg]<<1$. A careful examination
shows that all the m's may be approximated by $\mu$'s; in
particular, we also have $|\frac{\mu_l}{m_l}-1|\sim
O[\frac{m_m}{m_h}]$. Therefore (2.14) becomes
$$
\frac{4m_mm_l}{(m_m-m_l)^2}=\frac{1-r^2}{r^2}\sin^2B\eqno(2.22)
$$
and (1.7) and (1.8) are established.

Also, (1.9) is a direct consequence of (2.13) and (2.20). We may
take (1.10) as the definition of ${\cal L}$, and from (2.20) we may
write it as
$$
{\cal
L}=\frac{|L_\downarrow|}{m_b}-\frac{|L_\uparrow|}{m_t}.\eqno(2.23)
$$
The first equality of (1.11) is the definition of ${\cal S}$. Thus
what remains is to establish the second part of (1.11), and
(1.12)-(1.14). This requires studying the CKM matrix which relates
"up" to "down" eigenstates, as we shall see.

\section*{\Large \sf 3. The CKM Matrix}

In this section we restore the arrow subscripts
$\uparrow,~\downarrow$. On account of (2.16) and (2.21), the
matrix $W$ defined in (2.4) is given by
$$
W_{\uparrow,\downarrow}^\dag=P_{\uparrow,\downarrow}^\dag
V_{\uparrow,\downarrow}^\dag
\Omega_{\uparrow,\downarrow}^\dag.\eqno(3.1)
$$
If we define
$$
U=W^\dag_\uparrow W_\downarrow=P^\dag_\uparrow
U_0P_\downarrow\eqno(3.2)
$$
where
$$
U_0=V^\dag_\uparrow\Omega^\dag_\uparrow\Omega_\downarrow
V_\downarrow
$$
$$
=\left(
\begin{array}{c;{2pt/2pt}c}
\bigg(e^{\frac{1}{2}i\tau_yB_\uparrow}
e^{-\frac{1}{2}i\tau_zA_\uparrow}\bigg)&
\begin{array}{c}
0\\
0
\end{array}\\\hdashline[2pt/2pt]
0~~~~~~~~0&1
\end{array}\right)\left(
\begin{array}{c;{2pt/2pt}c}
1&0~~~~~0\\\hdashline[2pt/2pt]
\begin{array}{c}
0\\
0
\end{array}&
e^{i(\Phi_\uparrow-\Phi_\downarrow)\tau_y}
\end{array}\right)\left(
\begin{array}{c;{2pt/2pt}c}
\bigg(e^{\frac{1}{2}i\tau_zA_\downarrow}
e^{-\frac{1}{2}i\tau_yB_\downarrow}\bigg)&
\begin{array}{c}
0\\
0
\end{array}\\\hdashline[2pt/2pt]
0~~~~~~~~0&1
\end{array}\right)\eqno(3.3)
$$
then $U$ transforms eigenstates of $M_\downarrow$ into eigenstates
of $M_\uparrow$, provided that the phases of the eigenstates are
suitably chosen. To obtain the CKM matrix $U_{CKM}$, which relates
eigenstates whose phases follow a standard convention, we shall need
an additional transformation
$$
U_{CKM}=Q^\dag_\uparrow U Q_\downarrow\eqno(3.4)
$$
where $Q_{\uparrow,\downarrow}$ are diagonal unitary matrices to be
chosen presently.

In evaluating (3.3) it is convenient to introduce new symbols:
$$
\delta=\Delta_\uparrow\Delta_\downarrow^*=
e^{\frac{1}{2}i(A_\uparrow-A_\downarrow)},\eqno(3.5)
$$
$$
\Gamma=\cos\frac{1}{2}B_\uparrow,~~~\gamma
=\cos\frac{1}{2}B_\downarrow,\eqno(3.6)
$$
$$
\Sigma=\sin\frac{1}{2}B_\uparrow,
~~~\sigma=\sin\frac{1}{2}B_\downarrow,\eqno(3.7)
$$
$$
{\cal S}=\sin(\Phi_\uparrow-\Phi_\downarrow)~{\sf
and}~C=\cos(\Phi_\uparrow-\Phi_\downarrow).\eqno(3.8)
$$
We note that the first equation in (3.8) is the same in (1.11). By
using (3.5)-(3.8), we find $U_0$ of (3.3) can be written as
$$
U_0=\left(
\begin{array}{ccc}
\delta^*\Gamma\gamma+C\delta\Sigma\sigma&
-\delta^*\Gamma\sigma+C\delta\Sigma \gamma &{\cal
S}\Delta_\uparrow \Sigma\\
-\delta^*\Sigma \gamma+C\delta\Gamma\sigma& \delta^*\Sigma
\sigma+C\delta\Gamma \gamma& {\cal S}\Delta_\uparrow \Gamma\\
-{\cal S}\Delta_\downarrow^*\sigma &
 -{\cal S}\Delta_\downarrow^*\gamma &C
\end{array}\right).\eqno(3.9)
$$

The next step is to prepare for a perturbative treatment of (3.2)
by writing
$$
P_{\uparrow,\downarrow}\cong I+p_{\uparrow,\downarrow}\eqno(3.10)
$$
where (in arrowless notation)
$$
p^\dag=\frac{1}{m_h} \left(
\begin{array}{ccc}
0&0&-\Delta^*L\cos\frac{1}{2}B\\
0&0&\Delta^*L\sin\frac{1}{2}B\\
\Delta L^*\cos\frac{1}{2}B&-\Delta L^* \sin\frac{1}{2}B&0
\end{array}\right).\eqno(3.11)
$$
We note that by putting (3.11) into (3.10), we can satisfy (2.21)
to first order in $L$.

Thus we have
$$
U\cong U_0+U'\eqno(3.12)
$$
where
$$
U'=p^\dag_\uparrow U_0+U_0p_\downarrow.\eqno(3.13)
$$

Let us carefully evaluate the lower left element of
$p^\dag_\uparrow U_0$:
$$
(p^\dag_\uparrow U_0)_{31}=\frac{1}{m_t}(L^*_\uparrow
\Delta_\uparrow\cos\frac{1}{2}B_\uparrow)(\delta^* \Gamma
\gamma+C\delta\Sigma\sigma)
$$
$$
~~~~~~~~~~~~~~~~~~~~+\frac{1}{m_t}(-L^*_\uparrow
\Delta_\uparrow\sin\frac{1}{2}B_\uparrow)(-\delta^* \Sigma
\gamma+C\delta\Gamma\sigma)
$$
$$
~~~~~~~~~~~~~~~~~~~~~~~~=
\frac{L^*_\uparrow\Delta_\uparrow}{m_t}[\Gamma(\delta^*\Gamma
\gamma+C\delta\Sigma\sigma)+\Sigma(\delta^*\Sigma
\gamma-C\delta\Gamma\sigma)]
$$
$$
~~~~~~=\frac{L^*_\uparrow\Delta_\uparrow}{m_t}\delta^*
(\Gamma^2+\Sigma^2)\gamma
=\frac{L^*_\uparrow}{m_t}\Delta_\downarrow \gamma.\eqno(3.14)
$$
(Note how the calculation converts $\Delta_\uparrow$ to
$\Delta_\downarrow$ and $\Gamma$ to $\gamma$.) The corresponding
element of $U_0p_\downarrow$ is trivial:
$$
(U_0p_\downarrow)_{31}=C(\frac{1}{m_b}
\Delta_\downarrow^*L_\downarrow\cos\frac{1}{2}B_\downarrow)^*
=-\frac{L_\downarrow^*}{m_b}\Delta_\downarrow \gamma C.\eqno(3.15)
$$

Anticipating that $B_\uparrow$ will turn out fairly small, $\sim
0.2$, we now observe that the matrix element $U_{23}$ is going to
be dominated by $(U_0)_{23}={\cal S}\Delta_\uparrow\Gamma\sim
{\cal S}\Delta_\uparrow$. Therefore, ${\cal S}$ must have
magnitude $\sim .04$. It follows that $C\sim 1-\frac{1}{2}{\cal
S}^2$ can be replaced by $1$, and that all elements of $U'$ other
than $(U')_{13,23,31,32}$ being of order ${\cal
S}\cdot\frac{\sqrt{m_dm_s}}{m_b}$, can be neglected.

Thus, by repeating for $(U')_{13,23,32}$ the calculations leading
to (3.14) and (3.15), we have
$$
U'\cong \left(,
\begin{array}{ccc}
0&0&+(\frac{L_\downarrow}{m_b}-\frac{L_\uparrow}{m_t})\Delta_\uparrow \Gamma\\
0&0&-(\frac{L_\downarrow}{m_b}-\frac{L_\uparrow}{m_t})\Delta_\uparrow \Sigma\\
-(\frac{L^*_\downarrow}{m_b}-\frac{L^*_\uparrow}{m_t})\Delta_\downarrow
\gamma
&+(\frac{L^*_\downarrow}{m_b}-\frac{L^*_\uparrow}{m_t})\Delta_\downarrow\sigma
&0
\end{array}\right).\eqno(3.16)
$$
But from (2.8), taking $T,\beta~,~\cos\Phi,~\sin\Phi$ positive, we
find
$$
\frac{L_\downarrow}{|L_\downarrow|}=\frac{L_\uparrow}{|L_\uparrow}|
=\frac{1-e^{i\chi}} {|1-e^{i\chi}|}\eqno(3.17)
$$
and so
$$
\frac{L_\downarrow}{m_b}-\frac{L_\uparrow}{m_t}
=\frac{1-e^{i\chi}} {|1-e^{i\chi}|}{\cal L}\eqno(3.18)
$$
by (2.23). We now anticipate that $\chi$ will have to be negative
in order to make everything come out right. Hence,
$$
\frac{1-e^{i\chi}}
{|1-e^{i\chi}|}=\frac{e^{\frac{^1}{2}i\chi}(-2i\sin\frac{1}{2}\chi)}
{|2\sin\frac{1}{2}\chi|}=+ie^{\frac{1}{2}i\chi}\eqno(3.19)
$$
and (3.16) leads to
$$
U'\simeq \left(
\begin{array}{ccc}
0&0&+ie^{\frac{1}{2}i\chi}{\cal L}\Delta_\downarrow^* \Gamma\\
0&0&-ie^{\frac{1}{2}i\chi}{\cal L}\Delta_\downarrow^*\Sigma\\
+ie^{-\frac{1}{2}i\chi}{\cal L}\Delta_\downarrow \gamma
&-ie^{-\frac{1}{2}i\chi}{\cal L}\Delta_\downarrow\sigma &0
\end{array}\right).\eqno(3.20)
$$
For reasons shortly to be evident, let us now introduce the phase
factors
$$
\varepsilon_{\uparrow,\downarrow}=-i
e^{\frac{1}{2}i\chi}(\Delta^*_{\uparrow,\downarrow})^2=
e^{-\frac{i\sigma}{2}}e^{i(\frac{1}{2}\chi-
A_{\uparrow,\downarrow})}.\eqno(3.21)
$$
Then we have
$$
U'= \left(
\begin{array}{ccc}
0&0&-\varepsilon_\uparrow{\cal L}\Delta_\uparrow\Gamma\\
0&0&+\varepsilon_\uparrow{\cal L}\Delta_\uparrow\Sigma\\
+\varepsilon_\downarrow^*{\cal L}\Delta_\downarrow^* \gamma
&-\varepsilon_\downarrow^*{\cal L}\Delta_\downarrow^*\sigma &0
\end{array}\right).\eqno(3.22)
$$
In treating (3.9), let us note that since
$\Phi_\uparrow-\Phi_\downarrow\approx \sin^{-1}{\cal S}$ is small,
$A_\uparrow-A_\downarrow$ is also small by (2.10). Hence
$|Im~\delta|$ is small (see(3.5)) and $1-Re~\delta$ is second
order. So $Re~\delta$ can be taken $=1$, and the imaginary parts
of $(U_0)_{11,12,21,22}$ can be adjusted by small adjustments in
$Q_\uparrow, Q_\downarrow$. We shall treat such adjustments
imprecisely and simply neglect these imaginary parts. By taking
$C\rightarrow 1$ and using (3.6)-(3.7), we find
$$
\left(
\begin{array}{cc}
(U_0)_{11}&(U_0)_{12}\\
(U_0)_{21}&(U_0)_{22}
\end{array}\right)=
\left(
\begin{array}{cc}
\Gamma \gamma+\Sigma\sigma&-\Gamma \sigma+\Sigma \gamma\\
-\Sigma \gamma+\Gamma\sigma&\Sigma\sigma+\Gamma \gamma
\end{array}\right)
$$
$$
= \left(
\begin{array}{cc} \cos\frac{1}{2}(B_\downarrow-B_\uparrow)&
-\sin\frac{1}{2}(B_\downarrow-B_\uparrow)\\
\sin\frac{1}{2}(B_\downarrow-B_\uparrow)&
\cos\frac{1}{2}(B_\downarrow-B_\uparrow)
\end{array}\right).\eqno(3.23)
$$
Now $B_\downarrow-B_\uparrow$ must be positive to fit $U_{13}$ and
$U_{31}$, and so $U_{12}$ is negative, whereas the standard
presentation gives $(U_{CKM})_{12}$ positive. Therefore, we shall
use the $Q$-transformation to change the sign of the first row and
column, and also to remove the factors
$\Delta_\uparrow,~\Delta_\downarrow^*$ now appearing in the third
row and column. Thus
$$
Q^\dag_\uparrow=\left(
\begin{array}{ccc}
-1&0&0\\
0&1&0\\
0&0&\Delta_\downarrow
\end{array}\right),~~
Q^\dag_\downarrow=\left(
\begin{array}{ccc}
+1&0&0\\
0&1&0\\
0&0&\Delta_\uparrow^*
\end{array}\right)\eqno(3.24)
$$
and
$$
U_{CKM}=Q^\dag_\uparrow U_0Q_\downarrow+Q^\dag_\uparrow
U'Q_\downarrow
$$
$$
=\left(
\begin{array}{ccc}
\cos\frac{1}{2}(B_\downarrow-B_\uparrow)&
\sin\frac{1}{2}(B_\downarrow-B_\uparrow)&-{\cal
S}\Sigma+\varepsilon_\uparrow{\cal L}\Gamma\\
-\sin\frac{1}{2}(B_\downarrow-B_\uparrow)&
\cos\frac{1}{2}(B_\downarrow-B_\uparrow)& {\cal
S}\Gamma+\varepsilon_\uparrow{\cal L}\Sigma\\
{\cal S}\sigma-\varepsilon^*_\downarrow{\cal L} \gamma& -{\cal
S}\gamma-\varepsilon^*_\downarrow{\cal L}\sigma&1
\end{array}\right)\eqno(3.25)
$$
where we have again allowed a slight imprecision of phase in the
$(3,3)$ element.

Comparing (3.25) with the array
$$
U_{CKM} = \left(
\begin{array}{ccc}
(u|d)&(u|s) &(u|b)\\
(c|d)&(c|s) &(c|b)\\
(t|d)&(t|s) &(t|b)
\end{array}\right),\eqno(3.26)
$$
we obtain  the second half of (1.11) and (1.12)-(1.14).

Note: there is an ambiguity, $\Phi_{\uparrow,\downarrow}>~{\sf
or}~<\frac{\pi}{4}$. We take both $\Phi$'s$>\frac{\pi}{4}$, so
that $|A|>|\chi-A|$ and hence $|A|>|\frac{1}{2}\chi|$. Since
$\chi$ and $A$ are negative, $\frac{1}{2}\chi-A>0$ and hence
$Re~\varepsilon_{\uparrow,\downarrow}>0$, as required in $(u|b)$
and $(t|d)$. Because $Im~\varepsilon_\uparrow=
-\cos(\frac{1}{2}\chi-A)$, we can then derive (1.13) by using
(2.15).\\

\section*{\Large \sf 4. The "Timeon" Model}

The merit of the "strong $\gamma_4$ $T$-violation model" examined
in this paper suggests that there may be large $T$-violation
somewhere in physics although its manifestation in the quark mass
sector is small. In the "strong $\gamma_4$ $T$-violation model"
the $T$-violating effects are produced by the phase $\chi$ which
enters nonlinearly into the Hamiltonian. This non-linear
interaction makes it difficult to construct a renormalizable
quantum field theory that can be extended beyond the mass matrix.
For this and other reasons, we have considered a different
model[3] in which the $T$-violating effect enters linearly;
therefore, the model can lead to a renormalizable field theory,
called "timeon".

In the timeon theory, the mass-generating Hamiltonian can be
written by replacing $M_{\uparrow/\downarrow}$ in (1.4) by
$$
G_{\uparrow/\downarrow}+i\gamma_5
F_{\uparrow/\downarrow},\eqno(4.1)
$$
where $G_{\uparrow/\downarrow}$ and $F_{\uparrow/\downarrow}$ are
real symmetric matrices, and the $F_{\uparrow/\downarrow}$ term in
$i\gamma_4$ arises from coupling to the vacuum expectation value
of a new $T$-negative and $P$-negative field $\tau(x)$, the timeon
field. Thus, the whole field theory conserves $T$, but
$T$-violation arises from the spontaneous symmetry breaking that
makes the vacuum expectation value
$$
\tau_0=<\tau(x)>_{vac}\neq 0.\eqno(4.2)
$$

The timeon field $\tau(x)$ is real, so that there is no Goldstone
boson[4]. However, the oscillation of $\tau(x)$ around its vacuum
expectation value $\tau_0$ gives rise to a new particle, called
"timeon", whose production can lead to large $T$-violating effects.
In Ref.~3, it is shown that the parameters determining
$G_{\uparrow/\downarrow}$ and $F_{\uparrow/\downarrow}$ can be
adjusted to simulate an arbitrary complex $\gamma_4$ model, as far
as the quark masses are concerned, but not the CKM matrix. Thus, for
example, in the timeon $\gamma_5$-model the light quark masses in
the small mass limit turn out to be proportional to ${\cal J}$,
whereas in the $\gamma_4$-model, they are proportional to ${\cal
J}^2$.

\section*{\Large \sf References}

\noindent [1] R. Friedberg and T. D. Lee, Ann. Phys. {\bf 323}(2008)1087\\

\noindent [2] Particle Data Group, J. Phys. {\bf G33}(2006)1\\

\noindent [3] R. Friedberg and T. D. Lee, arXiv:0809.3633\\

\noindent [4] J. Goldstone, Nuovo Cimento {\bf 9}(1961)154\\

\vspace{1cm}

\begin{center}
{\Large \sf Table~1}

\vspace{1.cm}

\begin{tabular}{|c|c|}
  \hline
 Parameter & "Best" value\\
  \hline
 $m_s$ & $95MeV$ \\
% \hline
 $m_b$ & $4.5GeV$ \\
%\hline
 $(c|b)$ & $0.04$ \\
%  \hline
 $Re(u|b)$ & $0.002$ \\
%  \hline
 $Im(u|b)$ & $-0.003$ \\
  \hline
 \end{tabular}
\end{center}

 \vspace{1cm}

 These values are used to obtain the top two rows in Table~2.\\

\newpage

\begin{center}
{\Large \sf Table~2}\\

\vspace{.5cm}

 Values of $m_u,~m_d$ and $\chi$ calculated from the
strong $\gamma_4$-model

\vspace{1cm}

{\bf \normalsize
\begin{tabular}{|c|c|c|}
  \hline
 Input parameters && \begin{tabular}{ccc}
                   $m_u(MeV)$ & $m_d(MeV)$ & $\cos\frac{1}{2}\chi$
                   \end{tabular}\\
  \hline
  As in Table~1 &&  \begin{tabular}{ccc}
                   1.45 & ~~~~~~5.18~~~~~~ & .487\\
                   3.16 & 6.50 & .428
                   \end{tabular}\\
                   \hline
 Table~1 except & $m_s=85MeV$ & \begin{tabular}{ccc}
                   1.39 & ~~~~~~5.43~~~~~~ & .479\\
                   3.29 & 6.86 & .418
                   \end{tabular}\\
                   \hline
 Table~1 except & $m_s=105MeV$ & \begin{tabular}{ccc}
                   1.52 & ~~~~~~5.00~~~~~~ & .490\\
                   3.09 & 6.22 & .433
                   \end{tabular}\\
                   \hline
 Table~1 except & $m_b=4.2GeV$ & \begin{tabular}{ccc}
                   1.63 & ~~~~~~4.83~~~~~~ & .483\\
                   3.33 & 6.02 & .427
                   \end{tabular}\\
                   \hline
 Table~1 except & $m_b=4.7GeV$ & \begin{tabular}{ccc}
                   1.61 & ~~~~~~5.68~~~~~~ & .476\\
                   3.53 & 7.14 & .417
                   \end{tabular}\\
                   \hline
 Table~1 except & $(c|b)=0.035$ & \begin{tabular}{ccc}
                   1.40 & ~~~~~~4.86~~~~~~ & .507\\
                   2.98 & 5.96 & .454
                   \end{tabular}\\
                   \hline
 Table~1 except & $(c|b)=0.045$ & \begin{tabular}{ccc}
                   1.51 & ~~~~~~5.52~~~~~~ & .468\\
                   3.36 & 7.07 & .405
                   \end{tabular}\\
                   \hline
 Table~1 except & $Re(u|b)=0.0015$ & \begin{tabular}{ccc}
                   1.63 & ~~~~~~4.74~~~~~~ & .525\\
                   3.33 & 5.96 & .463
                   \end{tabular}\\
                   \hline
 Table~1 except & $Re(u|b)=0.0025$ & \begin{tabular}{ccc}
                   1.72 & ~~~~~~6.09~~~~~~ & .432\\
                   2.96 & 7.06 & .397
                   \end{tabular}\\
                   \hline
Table~1 except & $Im(u|b)=-0.0025$ & \begin{tabular}{ccc}
                   1.64 & ~~~~~~4.93~~~~~~ & .428\\
                   2.75 & 5.81 & .389
                   \end{tabular}\\
                   \hline
 Table~1 except & $Im(u|b)=-0.0035$ & \begin{tabular}{ccc}
                   1.73 & ~~~~~~5.96~~~~~~ & .510\\
                   2.93 & 6.83 & .473
                   \end{tabular}\\
                   \hline
 \end{tabular}
}

\end{center}

%\newpage

Table~2 (footnotes)\\

The values of five input parameters are taken as in Table~1,
except for single departures as shown in the left-hand column
here. For each setting of the input parameters, there is a
one-parameter family of solutions of Eqs. (1.7)-(1.14). We show
two members of each family, chosen roughly to span the
experimental range of $m_u$ from $1.5$ to $3.0MeV$. The
corresponding values of $m_d$ stay within its experimental range
from $3$ to $8MeV$, and $\chi$ remains large from $-120^0$ to
$-135^0$.\\

\vspace{2cm}

\centerline{\epsfig{file=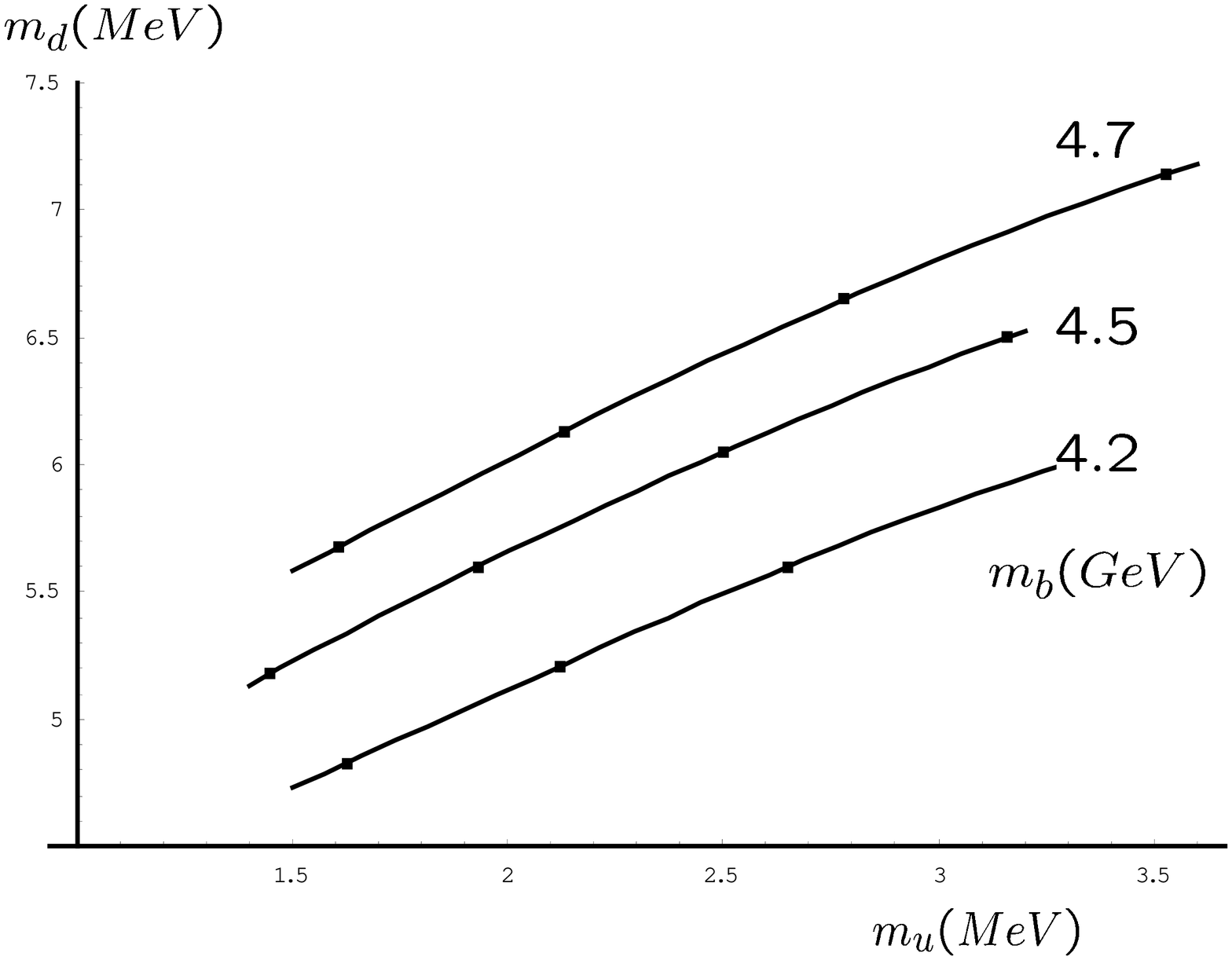,width=11cm,angle=270}}

\vspace{.5cm}

\noindent Figure~1. $m_d$ versus $m_u$ for $m_s=95MeV$,
$(c|b)=0.04$,

~~~~$(u|b)=0.002-0.003i$ and $m_b=4.2GeV,~4.5GeV$ and $4.7GeV$.

\end{document}